\documentclass[twocolumn,amssymb,showkeys,floatfix,12pt]{revtex4-1}
\usepackage{graphicx}
\usepackage{multirow}
\usepackage{float}
\usepackage{bm}
\usepackage{dcolumn} 
\usepackage{hyperref}
\begin{document} 
\title[How to understand the tunneling in attosecond experiment]
{How to understand the tunneling in attosecond experiment?
\\
{\scriptsize\sl Bohr-Einstein photon box {\it Gedanken experiment},
tunneling time and the wave particle duality}}
\author{Ossama Kullie}
\affiliation{Theoretical Physics, Institute for Physics, Department
of Mathematics and Natural Science, University of Kassel, Germany}
\thanks{\tiny Electronic mail: kullie@uni-kassel.de}
\begin{abstract} 
\scriptsize
The measurement of the tunneling time (T-time) in today's attosecond 
and strong field (low-frequency) experiments, despite its 
controversial discussion, offers a fruitful opportunity to understand  
time measurement and the time in quantum mechanics. 
In addition, as we will see in this work, a related controversial 
issue is the particulate nature of the radiation.  
Different models used to calculate the T-time will be 
discussed in this work in relation to my model of real 
T-time, Phys. Rev. {\bf 92}, 052118 (2015), where  an intriguing 
similarity to the Bohr-Einstein photon box {\it Gedanken experiment} 
was found. 
The tunneling process itself is still not well understood, but I am 
arguing that a scattering mechanism (by the laser wave packet) offers 
a possibility to understand the tunneling process in the tunneling 
region. 
This is related to the question about the corpuscular nature 
of light which is widely discussed in  modern quantum 
optics experiments. 
\end{abstract}  
\maketitle     
\scriptsize
\section{Introduction}\label{sec:In}
There is no doubt that the advent of 'attophysics' opens new 
perspectives in the study of time resolved phenomena in atomic and 
molecular physics 
\cite{Corkum:2008,Maquet:2014,Gallmann:2012,Kling:2008,Scrinzi:2006},  
the tunneling process and the tunneling time (T-time) in atoms and 
molecules 
\cite{Krausz:2009,Eckle:2008s,Eckle:2008,Landsman:2014II,Kullie:2015}. 
Attosecond science concerns primarily electronic motion and energy 
transport on atomic and molecular scales and is of fundamental interest 
to physics in general.
The time-energy uncertainty relation (TEUR) receives 
a ‘new breath’ due to the actual problems of quantum information 
theory and impressive progress of the experimental technique 
in quantum optics and atomic physics 
\cite{Dodonov:2015,Maquet:2014}.
In my previous work \cite{Kullie:2015}, I showed that using the TEUR 
(precisely that time and energy are conjugate variables) leads to a 
nice relation to determine the T-time in good agreement with the 
experimental finding in the attosecond experiment (for He atom) 
\cite{Landsman:2014II}, ($1\,\mbox{attosecond}=10^{-18}$ second). 
The T-time and time itself in quantum mechanics (QM) are 
controversial, and there is still common opinion that time plays 
a role essentially different from the role of position in quantum 
mechanics (although it is not in line with special relativity,
\cite{BauerM:2014}) and that time is  a parameter, like a classical 
Newtonian time quantity, and hence does not obey an ordinary TEUR. 
Nevertheless, Hilgevoord concluded in his work \cite{Hilgevoord:2002} 
that when looking to a time operator a distinction must be made 
between the universal time coordinate $t$, a c-number like a space 
coordinate, and the dynamical time variable of a physical 
system situated in space-time; i.e. clocks. 
Accordingly in \cite{Kullie:2015,Kullie:2016} it was shown that the 
T-time is intrinsic, i.e. dynamically connected to the system 
(internal clock) after the classification of Busch 
\cite{Busch:1990I,Busch:1990II} and \cite{Muga:2008} (chap. 3). 
Fortunately, Bauer's introduction of a self-adjoint dynamical 
time operator in Dirac’s relativistic quantum theory  
\cite{BauerM:2014,BauerM:2016}, supports the results of 
\cite{Kullie:2015}. 
In \cite{BauerM:2016,BauerM:2016I} Bauer concluded that the dynamical 
time operator provides a straightforward explanation (within standard 
relativistic quantum mechanics) of the T-times, which is measured in 
the photoionization experiments, compare the discussion 
in \cite{Kullie:2016}.
In this respect, Bauer also rejects the claim of Dodonov 
\cite{Dodonov:2015} that no unambiguous and generally accepted results 
have been obtained for the time operator 
\cite{BauerM:2016,BauerM:2016p}.  
Moreover, Bauer showed \cite{BauerM:2016} that the Mandelstam-Tamm 
uncertainty associated with the observable $\hat{T}$ largely 
overestimates the internal time standard uncertainty as already 
discussed by Kullie \cite{Kullie:2016}. 

A similar controversial issue to the time issue (and the T-time and 
TEUR) is the wave-matter duality and the particulate nature of the 
light \cite{Lamb:1995,Zeilinger:2005}, since the Einstein hypothesis 
of the quanta as a carrier of $h\nu$ based on the Planck 
hypothesis of the quantization of the energy $E=h\nu$. 
The term photon was given by G. N. Lewis in 1926 \cite{Lewis:1926}, 
and indeed the corpuscular hypothesis originally stems from Newton. 
As we will see in this work, the two issues closely appear in 
today's attosecond experiments (ASEs). 
Indeed, since the appearance of QM time  was controversial, 
the famous example is the Bohr-Einstein weighing 
{\it photon box Gedanken experiment (BE-photon-box-GE)}.
In \cite{Kullie:2015} I showed with a simple tunneling model that the  
tunneling in the attosecond experiment is intriguingly similar to the  
{\it BE-photon-box-GE}, where the former can be seen as a realization 
to the later, with the electron as a particle (instead of the photon) 
and an uncertainty in the energy being determined from the (Coulomb) 
atomic potential due to the electron being disturbed by the field $F$, 
instead of (the photon) being disturbed by the weighting process and, 
as a result, an uncertainty in the gravitational potential 
\cite{Aharonov:2000}, as shown by the famous proof of Bohr 
(see for example \cite{Auletta:2009} p. 132) to the uncertainty 
(or indeterminacy) of time in the {\it BE-photon-box-GE} 
\cite{Aharonov:2000,Busch:1990I,Busch:1990II}.

\begin{figure}[t]
\includegraphics[height=4.0cm,width=6.0cm]{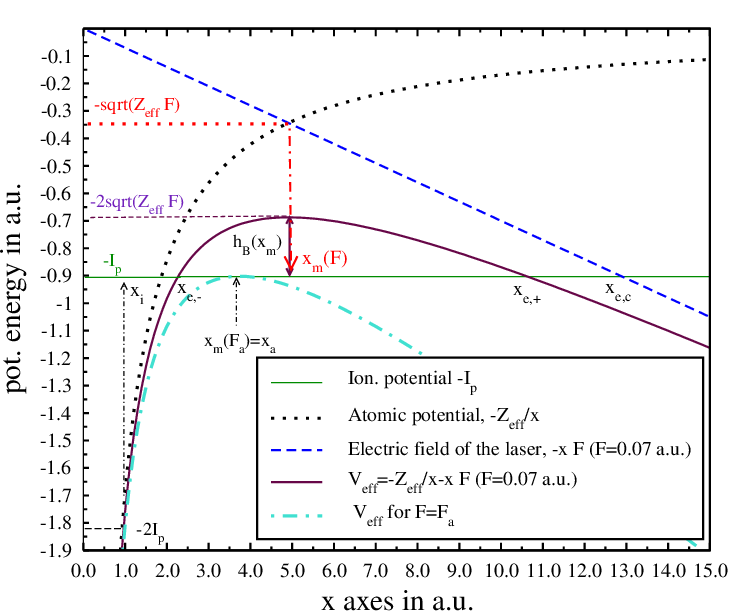}
\caption{\label{fig:ptc}\tiny (Color online)
Graphic display of the potential and the effective potential curves,  
the two inner and outer points $x_{e,\pm}={(I_p\pm\delta_z)/2F}$ 
and the barrier width $d_B=x_{e,+}-x_{e,-}$, the ``classical exit'' 
point $x_{e,c}=I_p/F$ and $x_m(F)=(Z_{_{eff}/F})^{1/2}$  the 
position at the maximum of the barrier height $h_B(x_m)$, 
(note $x_a=x_m(F=F_a)$), see text.}
\vspace{-0.50cm}
\end{figure}

The T-time and the tunneling process itself in the ASEs are hot 
debated, and the later is still rather unresolved puzzle.
In the (low-frequency) ASEs the idea is to control the 
electronic motion by laser fields that are comparable in strength 
to the electric field in the atom. In today's experiments usual 
intensities are $\sim 10^{14}\, W cm^{-2}$, for more details we refer 
to the tutorial 
\cite{Calegari:2016,Dahlstrom:2012,Krausz:2009,Kling:2008,Scrinzi:2006}. 
In the majority of phenomena in attosecond physics, one can separate 
the dynamics into a domain ``inside'' the atom, where atomic forces 
dominate, and ``outside'', where the laser force dominates, a two-step 
semi-classical model, pioneered by Corkum \cite{Corkum:1993}.
Ionization as the transition from ``inside'' to ``outside'' of the 
atom plays a key role for attosecond phenomena. A key quantity is 
the Keldysh parameter \cite{Keldysh:1964},
\begin{equation}\label{gamK}
\gamma_{_K} = \frac{\sqrt{2I_p}}{F} \omega_0=\tau_K\,  \omega_0, 
\end{equation} 
where $I_p$ denotes the ionization potential of the system (atom or 
molecule), $\omega_0$ is the central circular frequency of the laser 
pulse or the laser wave packet (LWP) and $F$, throughout this work,  
stands for the peak electric field strength at maximum, and $\tau_K$ 
denotes the Keldysh time. 
Hereafter in this work (unless it is clear), atomic units are used, 
where $\hbar=m=e=1$, the Planck constant, the electron mass and the 
unit charge are all set to 1.
At values $\gamma_{_K}>1$ one expects predominantly photo-ionization 
or multiphoton ionization (MPI), while at $\gamma_{_K}<1$ 
(field-)ionization happens by a tunneling process (for $F<F_{a}$), 
which means that the electron does not have enough energy to ionize 
directly, and therefore it tunnels (or tunnel-ionizes) through 
a barrier made by the Coulomb potential and the electric field of the 
laser pulse and escapes at the exit point to the continuum, as shown 
in fig \ref{fig:ptc}, see the following section.
We pay  attention to one important case study in 
attosecond physics, the T-time measurement in ASE performed by Keller  
\cite{Eckle:2008s,Eckle:2008,Landsman:2014II} and we will refer to it 
as the Keller ASE (KASE).  
In this experiment an elliptically polarized laser pulse is used with  
$\omega_0=0.0619\,au\,(\lambda=736\, nm)$, the ellipticity parameter 
$\epsilon=0.87$, while the electric field strengths are in the 
range $F=0.04-0.11$ and for He atom $I_p=0.90357\, au$.
\section{The tunneling time}\label{sec:tt}
Usually the tunneling process in the ASEs is explained by a simple 
picture, like the one shown in fig \ref{fig:ptc} for the He atom. 
It is based on the strong field approximation (SFA) or 
Keldysh-Faisal-Reiss approximation 
\cite{Keldysh:1964,Faisal:1973,Reiss:1980}; for an introductory review 
see \cite{Ivanov:2005}. 
This simple picture is very useful in explaining the experiment, 
although it is strictly true only in length gauge 
\cite{Bauer:2005,Faisal:2007I,Faisal:2007II}. 
Physical quantities are independent of a gauge transformation as 
long as exact equations (or the same orders of approximation 
\cite{Fried:1973})  are employed.  
Indeed, the length gauge or the dipole approximation of the 
interaction Hamiltonian due to the G\"oppert-Mayer 
gauge-transformation (when long-wavelength approximation applied) also 
has the advantage that it leads to an expression for the interaction 
energy involving mathematical quantities, each of which has a ready 
physical interpretation \cite{Grynberg:2010}. 

In the tunneling process in the ASEs, according to the SFA (see also 
\cite{Perelomov:1966}), the electron tunnels and escapes the barrier 
region at the exit point $x_{e,+}$, see fig \ref{fig:ptc}, with 
approximately zero kinetic energy, more precisely the electron 
velocity along the field direction is zero and negligible in the 
other directions. 
In \cite{Kullie:2015} (hereafter Kullie model) we showed that the 
uncertainty in the energy, which is related to the hight of the 
barrier $h_B(x_m)$, can be quantitatively discerned from the atomic 
potential energy at the exit point 
$\Delta E \sim \mid\! V(x_e)\!\mid=\mid\!-{Z_{eff}}/{x_e}\!\mid$
for arbitrary strengths $F\le F_a$, where $Z_{eff}$ is the nuclear 
effective charge and $F_a=I_p^2/(4 Z_{eff})$ is the atomic field 
strength \cite{Augst:1989,Augst:1991} 
\footnote[1]{\tiny I have to express that when I started my work 
\cite{Kullie:2015}, I was inspired by a talk of J. H. Eberly presented 
at a workshop on atomic physics, which was organized by Wilhelm and 
Else Heraeus-Stiftung in Germany, www.we-heraeus-stiftung.de}.  
With the TEUR, $\Delta E \cdot \Delta T \le 1/2$, one obtains 
the symmetrical (or total) T-time \cite{Kullie:2015}: 
\begin{eqnarray}\label{Tsym}
\hspace*{-0.5cm}\tau_{_T,sym}&=&
\tau_{_{T,i}}+\tau_{_{T,d}}
=\frac{1}{2}\left(\frac{1}{(Ip+\delta_z)}
+\frac{1}{(Ip-\delta_z)}\right), 
\end{eqnarray}
where $\delta_z = \sqrt{I_p^2-4 Z_{eff} F}$. 
The relation in eq \ref{Tsym}, besides the mathematical simplicity, 
aids a conceptual reasoning \cite{Kullie:2015,Kullie:2016} and the 
discussion further below.
The physical reasoning of this relation is the following: 
the barrier itself causes a delaying time $\tau_{_{T,d}}$, which 
is the time delay with respect to the ionization at atomic field 
strength $F_a$, where the barrier is absent (i.e. the barrier height, 
the barrier width and $\delta_z$ are zero), it is  the time duration 
to pass the barrier region (between $x_{{_e,-}},  x_{_{e,+}}$) and 
escape at the exit point $x_{e,+}$ to the continuum, for more details 
see \cite{Kullie:2015}. 
The first term $\tau_{_{T,i}}$ in eq \ref{Tsym} is the time needed 
to  reach the entrance point $x_{e,-}$ from the initial point $x_i$, 
compare fig \ref{fig:ptc}.
At the limit $F\rightarrow F_a$, $\delta_z\rightarrow 0$ and the total 
time becomes the ionization time $\tau_{_T,sym}=\frac{1}{Ip}$ at the 
atomic field strength $F_a$. 

The T-time is a controversial issue from different points of view, 
as we will discuss in the following subsections.  
\subsection{\small{A real or an imaginary quantity}}\label{ssec:riq}
\begin{figure}[t]
\includegraphics[height=4.0cm,width=6.0cm]{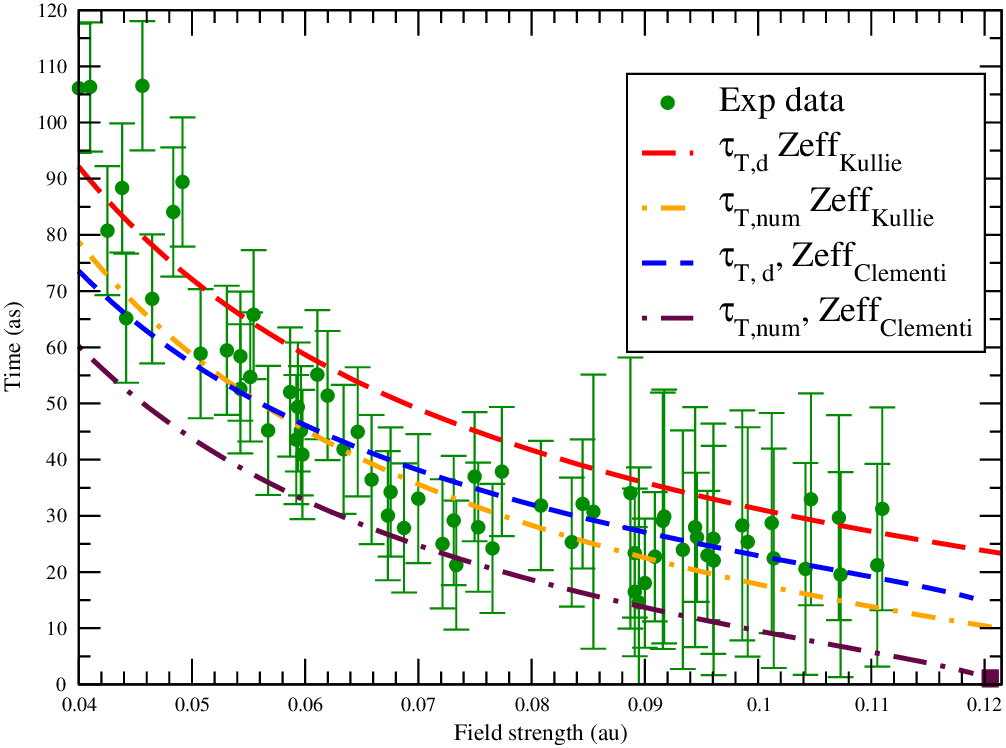}
\caption{\label{fig:dnum1}\tiny (Color online)
Graphic display of tunneling time  vertical axes (in attosecond) 
vs field strength $F$ in $au$. $\tau_{T,d}$ eq \ref{Tsym} (red 
$Zeff_{_{Kullie}}$, blue $Zeff_{_{Clemanti}}$, and $\tau_{num}$ 
eq \ref{Tnum} (orange $Zeff_{_{Kullie}}$, brown $Zeff_{_{Clemanti}}$). 
$\tau_{T,d}, \tau_{num}$  are plotted with $F$ as a free variable. 
Experimental values from \cite{Landsman:2014II}.}
\end{figure}

First, the most reasonable argument is that that T-time is a real 
quantity, as clarified by Steinberg \cite{Muga:2008} (chap. 11) 
and B\"uttiker  \cite{Muga:2008} (chap. 9), although  many authors 
claim, and it is widely accepted, that it is an imaginary quantity 
\cite{Torlina:2015,Shvetsov:2016}, \cite{Sokolovski:1990}  
(and \cite{Muga:2008} chap. 7).   
The imaginary tunneling time point of view relies on the fact that 
the tunneling is classically forbidden. 
Although that is important, it is not instructive, in the case it 
obscures an insight and otherwise accessible conceptual understanding 
\footnote[2]{\tiny See Carver Mead, the Nature Of Light What Are Photons, 
 www.cns.caltech.edu}.
One has to mention that a complex time point of view 
(i.e. real and imaginary parts) would not change the conclusion 
of our works, because the crucial point in our discussion is a real 
part of the T-time. 
In agreement with the real T-time point of view, with the real time of 
the FPI description of \cite{Landsman:2014II} (although Sokolovski 
\cite{Sokolovski:1990},  \cite{Muga:2008} chap. 7, argued that T-time 
described by the FPI is imaginary), and the entropic formulation 
(or the statistical approach) of (real) T-time of Demir 
\cite{Demir:2016}, the relation in eq \ref{Tsym} presents a real 
T-time  model which explains the T-time in KASE in a good agreement 
with the experimental finding \cite{Kullie:2015,Kullie:2016}, 
compare fig \ref{fig:dnum1}, \ref{fig:dnum2}.  
Although the model is simple, it is important in the quest to answer 
the question: how to understand the T-time and tunneling process, the 
time measurement and the time in quantum mechanics \cite{Kullie:2016}.  

Secondly, the treatment in \cite{Kullie:2015,Kullie:2016} benefits 
from  the internal (intrinsic or dynamical) time point of view 
(internal clock, \cite{Muga:2008} chap. 3), this requires one to 
choose a ‘reference’ point 
\cite{Maquet:2014,Hilgevoord:2002,Aharonov:2000}, which can be at best 
determined by the (natural) internal properties of quantum mechanical 
systems (e.g. ionization potential $I_p$). 
This is important as it enables one to identify or map the internal 
delay time (a time interval) as a delay time in an external clock 
without, certainly, contraction or dilation of the time (scale unity) 
interval of the clocks 
(i.e. no effects like in the relativistic theory.)  
We also note that the internal clock or the intrinsic time point of 
view is similar to what occurs in special relativity, where a moving 
particle has its own time in its inertial frame, which differs from 
the time from the viewpoint of other inertial frames, as discussed 
in \cite{Kullie:2016}.
Finally we mention that some authors \cite{Torlina:2015,Shvetsov:2016} 
use the notation $t_s=t_R+ i\, t_T$, where $t_s$ refers to the 
solution of the saddle point equation, $t_R$ or the real part of $t_s$ 
denoted as the ionization time (after tunneling) and the imaginary part 
$t_T$  as the T-time.
The partition of $t_s$ this way, in a real part for ionization and an 
imaginary part for the tunneling, lets some questions be opened.  
We discus this in the next subsec \ref{ssec:mda}.   
It is worthwhile to mention (see discussion in \cite{Kullie:2015}) 
that in eq \ref{Tsym}, $\delta_z$ becomes imaginary 
$\delta_{im,z}=i\,\delta_z$ ($i$ the imaginary number) for 
$F \ge F_a$ and that is the above-threshold 
region. 
\subsection{\small{Many different approaches}}\label{ssec:mda} 
The scattering theory concept is widely used to  calculate the 
T-time from the scattering time (for example B\"uttiker-Landauer or 
Pollak-Miller time, for details see \cite{Collins:1987,Landsman:2015}. 
However, following Collins \cite{Collins:1987} this is not justified. 
Recently Landsman \cite{Landsman:2014II} showed that these time 
approximations are in disagreement with the experimental 
finding of KASE, and that the FPI with a coarse graining procedure 
fits well with the KASE measurement data. 
Collins's most critical point \cite{Collins:1987}, which addresses the 
question: how long does it take for an electron to tunnel through a 
barrier, was the conclusion that the scattering time, defined through 
the inverse of the transition probability matrix, is not related to 
the dynamical transport behavior of the electron's tunneling. 
It is associated with the finite lifetime and decay of metastable 
states, in this case being the tunneling electrons treated 
as quasi-particles which are decaying from a state on one side of 
the potential barrier into another state on the opposite side of 
the barrier.  
Since this quantifies the decay of metastable states and not the 
transit time of an electron across a barrier, this time can be quite 
large and is a steady-state picture which does not reflect the 
dynamical nature of the tunneling particle(s).  
And according to Collins, 
it can be seen on the basis of the time-independent picture and by 
analogy with a particle decay, that the scattering time represents 
a mean time in  which a certain likelihood of a tunneling event 
may take place.  
This does not reflect the actual time of a tunneling 
process \cite{Collins:1987}. 
A similar point of view was considered by Fock and Krylov 
\cite{Fock:1947}, in regards to the lifetime and TEUR, see 
discussion by Aharonov \cite{Aharonov:1961}. 
\begin{figure}[t]
\includegraphics[height=4.0cm,width=6.0cm]{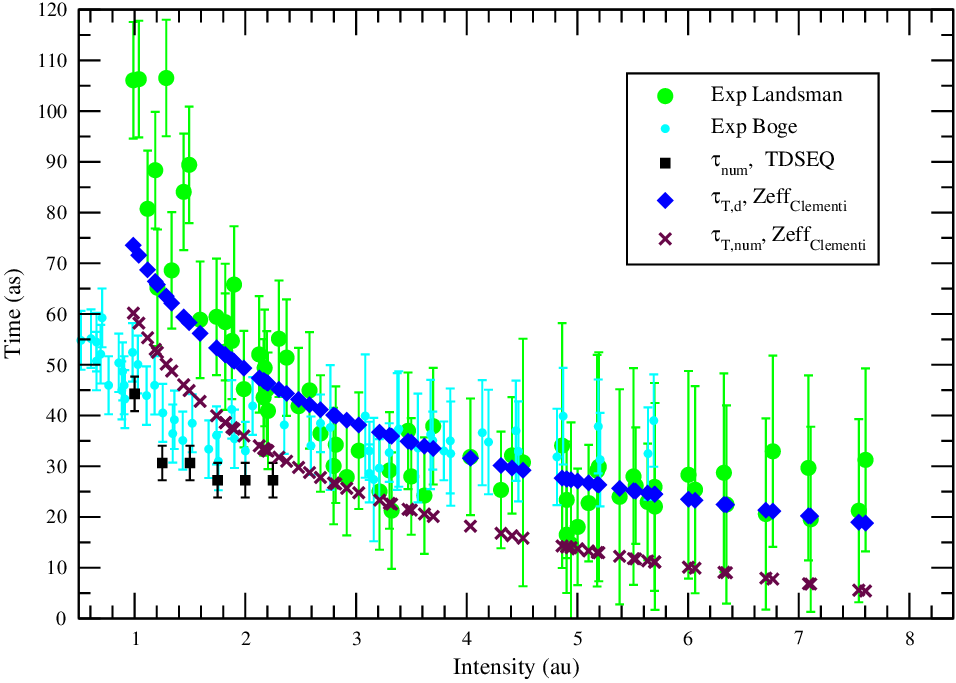}
\caption{\label{fig:dnum2}\tiny (Color online)
Graphic display of tunneling time vertical axes (in attosecond) vs
the intensity in $au$.  
$\tau_{T,d}, \tau_{num}$ are plotted at the experimental value of 
\cite{Landsman:2014II}. $\tau_{T,d}$  (upper, blue) eq \ref{Tsym}, 
and $\tau_{num}$ (lower, brown) eq \ref{Tnum} with 
$Zeff_{_{Clemanti}}$, see text.
The black squares are numerical integration of TDSEQ form 
\cite{IAIvanov:2014}, data kindly sent by I. A. Ivanov.
The points with error bars are experimental data, green 
(large circles) from Landsman \cite{Landsman:2014II} and the light 
blue (small circles) from R. Boge \cite{Boge:2013}, data kindly sent 
by A. S. Landsman and R. Boge.}
\vspace{-0.50cm}
\end{figure}

Collins then showed (using a Gaussian wave packet in the simulation), 
that the phase-time is overall the best one to use when momentum 
skewing of the initial wave packet is not significant.   
Unfortunately, in \cite{Kullie:2016}  we showed first that the phase 
time in attosecond experiment leads to the Keldysh time $\tau_K$, 
eq \ref{gamK} (an approximation that neglects the effect of the core 
potential, and could be important for the evolution of the wave packet), 
which is far from the experimental finding \cite{Landsman:2014II}. 
And second that a time-delay requires us to choose a reference 
(point);  delays in numerical simulation can refer in principle to any 
arbitrarily chosen reference \cite{Maquet:2014,Kullie:2016}. 
Indeed, because $\tau_{T,d}$ in  eq \ref{Tsym} presents a delay 
time relative to the limit at  atomic field strength 
$\tau_{T,d}(F=F_a)=1/(2I_p)\equiv \tau_{T,d,F_a}$, by subtracting 
the latter from former we get
\begin{eqnarray}\label{Tnum}
\hspace{-0.1cm}\tau_{T,d} - \tau_{T,d,F_a}&=& \frac{1}{2(I_p-\delta_z)}-
\frac{1}{2I_p}=
\frac{(\frac{\delta_z}{I_p})}{2(I_p-\delta_z)}\equiv
\tau_{num}\\\nonumber
&=&\frac{\delta}{2Ip^2}+
\cdots \frac{\delta^k}{2Ip^{k+1}} \cdots=\frac{1}{2Ip}\sum_{k=1}
\left(\frac{\delta}{Ip}\right)^k 
\end{eqnarray}
where in the second line an expansion of the form 
$(1-x)^{-1}= 1+x+x^2 +..., \,\, x=(\delta/Ip)$, is used, and 
$1/(2I_p)$ is the zero order term ($k=0)$). 
In fig \ref{fig:dnum1} the T-time $\tau_{T,d}$ of eq  \ref{Tsym} 
and $\tau_{num}$ of eq \ref{Tnum} are compared with experimental data of 
\cite{Landsman:2014II}, the field strength $F$ is the free variable and 
runs between $0.04-0.12$, the small dark square (at the right lower 
corner) marks $F_a$, where $\tau_{num}$  becomes zero and 
$\tau_{T,d}={1}/{2Ip}$ as already mentioned.
The crucial difference between  $\tau_{num}$ and $\tau_{T,d}$ is that 
the former tends to zero at atomic field strength ($\delta_z(F_a)=0$), 
$\lim_{F\to F_a} \tau_{num}=0$, which can happen only numerically, 
whereas quantum mechanically 
$\lim_{F\to F_a} \tau_{T,d}=\frac{1}{2I_p}$ is the second part of 
the total ionization time, eq \ref{Tsym}, at  atomic field strength 
($\tau_{T,sysm}=\frac{1}{2I_p}+\frac{1}{2I_p}=\frac{1}{I_p}$), 
as discussed in \cite{Kullie:2015}. 
This can be seen in fig \ref{fig:dnum1} for $Zeff=1.6875$ of Clementi 
\cite{Clementi:1963}, where $F_a \approx 0.12\, au$, which has to be 
chosen in this region and matches the experiment.
$Zeff=1.375$ of Kullie ($F_a\approx 0.14\ au$) \cite{Kullie:1997} is  
the better choice in the region for small field strengths, for 
detailed  discussion see  \cite{Kullie:2015}. 

In his work \cite{BauerM:2016I}, titted the problem of time in quantum  
mechanics, Bauer mentioned that time interred in the time dependent 
Schr\"odinger equation (TDSEQ) has to be identified with a parameter 
(parametric time $t_{parm}$) that corresponds to the time coordinate of 
the laboratory frame of reference as claimed by Briggs 
\cite{Briggs:2008}. 
Although $\tau_{num}$ differs qualitatively from a $\tau_{parm}$ 
(defined below), because it is constructed form the dynamical 
(internal) time of the systems, it looks to be identical with 
the parametric time $\tau_{parm}$, 
i.e. one can write   
\begin{equation}\label{Tnump}
\tau_{parm}=\tau_{num}=t_{T,d}-\frac{1}{2Ip}, 
\end{equation} 
where $\tau_{parm}$ denotes a time interval such that 
$\tau_{T,d}=\tau_{parm}+1/(2I_p)$ and $1/(2I_p)$ is the ionization 
time (at atomic field strength $F_a$).
Thus, the identification of the time in the TDSEQ as a parametric 
time indicates the lack of a proper mapping of the time from the 
internal (time-frame) to an external time (time-frame) and not an 
inherent property of the time in TDSEQ i.e. in quantum mechanics, 
in agreement with Hilgevoord point of view and Bush classification 
as mentioned in sec.\ref{sec:In}.   
Thus; so $\tau_{num}$ is identical with the time interring the 
TDSEQ, when mapping the dynamical (internal) time to an external 
(laboratory) time by ignoring the reference point (i.e. neglecting 
$1/(2I_p)$), as clearly seen in eq \ref{Tnump} (and in figs 
\ref{fig:dnum1}, \ref{fig:dnum2}, see below.)   
Whereas $\tau_{T,d}$ counts the internal time (interval) and transforms 
it to the external time (interval, measurement data) when the electron  
moves from inner to the outer region (the tunneling or ionization 
process) due to the interaction with the laser field. 
Contrary to the (parametric) time of measurement thought by Aharonov  
\cite{Aharonov:1961}, that the time of measurement belongs to the 
observing apparatus, which has been corrected in his recent work 
\cite{Aharonov:2000}.
Note a time interval refers to a clock, which measures the time 
quantity with respect to a reference point, whereas a time variable 
denotes a time quantity of any type, see subsec \ref{ssec:riq} 
2nd part. 
And as discussed in \cite{Kullie:2016} perhaps only the classical 
Newtonian time is a parametric (external non-dynamical) type of time.
In fig \ref{fig:dnum2} the experimental data of \cite{Boge:2013} 
(light blue) and \cite{Landsman:2014II} (green) are displayed, they 
are from the same set of experimental data, but a renewed calibration 
procedure of the field strength was used in the later. 
One sees that the T-time points resulting from numerical integration of 
the TDSEQ of Ivanov \cite{IAIvanov:2014} (black squares), lie somehow 
below the experimental data of \cite{Boge:2013}, after shifting the 
experimental data (done by Landsman) one sees that our $\tau_{num}$ eq  
\ref{Tnum}, where the inserted $F$ values are taken from the experiment, 
lies below the experimental data of Landsman \cite{Landsman:2014II} in 
a similar way the points of Ivanov \cite{IAIvanov:2014} lie below the 
experimental data of Boge \cite{Boge:2013} \footnote[3]{\tiny Actually 
for the data of Landsman \cite{Landsman:2014II}, the conversion of the 
new calibrated field strengths results in a new set of data on 
intensity-scale as seen in the figure.}. 

Nevertheless, Collins conclusion shows that the usual scattering 
concepts when used for the ASEs are best suited to the MPI region, 
where momentum skewing of the electronic wave packet is small 
\cite{Collins:1987}, compare subsec \ref{ssec:dr}. 
Therefore, it is not surprising that Torlina el al, when using ARM 
(analytical R-matrix theory) to calculate the T-time in the ASEs 
(the attoclock), argued that the ARM  requires sufficiently thick 
tunneling barriers \cite{Torlina:2015,Torlina:2012I,Torlina:2012II}.   
Torlina el al calculated the T-time of the (field-)ionized electron 
from the ground state of the Hydrogen atom, which is exposed to 
an attosecond laser pulse \cite{Torlina:2015}. 
They also claim that T-time is imaginary (see \ref{ssec:riq}) and no 
real tunneling delay time is associated with the tunnel-ionization 
\cite{Torlina:2015}, i.e. for $F<F_a$, and optical tunneling 
is instantaneous. 
Instantaneous here means that a real T-time is zero, although it 
should, in fact, represent the time of a real dynamical process  
and that at the tunnel exit the electron is far from the nucleus, 
i.e. the tunnel distance is not negligible as confirmed 
experimentally by \cite{Hickstein:2012}. 
However,  it can be easily shown that the request of sufficiently 
thick tunneling barriers corresponds to the region of small field 
strength, for which (at optical frequencies) most likely 
$\gamma_K\sim \frac{1}{F}>1$, see fig \ref{fig:ptc} and eq \ref{gamK}, 
and that is the MPI region, where tunneling is not probable.  
In addition, it could be that the imaginary T-time of 
Torilina et al is related to the phase time of Collins 
\cite{Collins:1987} as discussed above, see also \cite{Kullie:2016}. 
In this case one expects that the phase time of the electronic wave 
packet, like the phase velocity, has no real physical significance, 
although it can be viewed as a characteristic of the evolution of the 
wave packet in the sense described by Messiah \cite{Messiah:1961}, 
see discussion in \cite{Kullie:2016}. 
\vspace{-0.5cm}
\subsubsection{In depth discussion}
\vspace{-0.3cm}
{
\tiny
A point, which I think  is important here, is that  the calculations 
of Torlina el al should be compared with the 
time-of-arrival (\cite{Muga:2008} chap 3, 
\cite{Allcock:1969I,Allcock:1969II}), 
\footnote[4]{\tiny For example, the time of arrival of the decay 
products at a detector, 
see \cite{Muga:2008} chap 10}.  
The later concerns the subsequent propagation of the tunnel-ionized 
electron (see below). One notes that in attosecond and strong-field 
science the (tunnel-)ionization process cannot be understood in its 
conventional form known in quantum optics \cite{Nubbemeyer:2008}. 
Important here (compare \cite{Torlina:2015}): 
First the calculation of the ionization time of Torlina et al 
(hereafter Torlina model) relies on the dynamic interaction of the 
outgoing electron with the remaining core, i.e. the effects of the 
long-range potential (starting at the exit point), and that the 
ionization is not yet completed at the moment the electron exits 
the tunneling barrier (note the different definitions of the barrier, 
see below).  
This looks similar to the concept of the time-of-arrival  method and 
the observable-type of time after the classification of Busch 
(\cite{Muga:2008} chap 3, where in this case the atom acts as a source
(confining the electronic wave packet up to $Re(t)=0$) 
\cite{Muga:2008} chap. 1, and where the real time counting, as Torlina 
et al also argued, starts at the tunnel exit of their model (which, 
however, defines the border between inner and outer regions.) 
Indeed, the potential at the exit point $x_{exit}=x_{e,+}$ defines 
the energy uncertainty in the model of Kullie \cite{Kullie:2015}, here 
we indicate a common point (consideration) between the two models, 
and that the potential energy at the exit point is the central 
quantity, although the exit points $x_{e,+}$ and $x_{e,c}$, and the 
assignment of the inner and outer regions in the tunneling process are 
different in the two models, see fig \ref{fig:ptc}.
We argue that, the electron escapes the tunnel exit $x_{e,+}$ 
(where the energy uncertainty is defined), $x_{e,+}>x_{e,-}>x_i$ 
($x_i$ is the initial position, see fig \ref{fig:ptc}) is real and 
hence it is reached by a real time, and in the outer region (after 
$x_{e,+}$, where according the SFA the electron is free) the effect 
of the long range potential decreases rapidly, i.e. the measurement 
data of \cite{Landsman:2014II} corresponds to the crossing of the 
barrier region (between $x_{e,-}$ and $x_{e,+}$) when the tunneling 
process occurs \cite{Landsman:2014II,Kullie:2015}, in the length 
gauge picture. 
Our real T-time $\tau_{T,d}$  represents a delay time with respect to 
the natural reference point (i.e. ionization at the atomic field 
strength $F_a$), which is certainly real, since the exit point 
$x_{e,+}\ge x_{e,-}> x_i$ (equality for $F_a$) is real, and quantum 
mechanically such a tunneling/ionization dynamical time can not be 
zero or purely imaginary. 

And second, the authors of \cite{Torlina:2015} claim that although the 
measured quantity (the electron momentum) is real, the trajectories in 
the ARM method are not classical, in the sense that the trajectories 
have both real and imaginary components all the way to the detector,  
where they claim that the real part of the trajectory starts 
near the origin without an explicit definition \cite{Torlina:2013}.      
This, in turn, shows that in the tunneling process, real and 
imaginary components of a trajectory and (hence) real and 
imaginary components of time can exist (quantum mechanically) in both 
the inside and outside regions (under the barrier and after tunneling,    
despite the differences in the definition of these regions). 
Thus the barrier region is not necessary captured solely by an 
imaginary time component although it is classically forbidden.

This leads to the conclusion, that the partition $t_s=t_R+i\, t_T$ 
mentioned above in subsec \ref{ssec:riq}, i.e. an imaginary part $t_T$ 
for tunneling and a real part $t_R$ for ionization after the tunneling, 
is at least not unique if acceptable, apart form the fact that time 
delay requires one to choose a reference system 
\cite{Dahlstrom:2012,Kullie:2016}. 
This partition is only in line with the classical point of view that 
the dynamics under the barrier is classically forbidden, hence no 
real T-time can exist for the dynamics in this region, i.e. the 
tunneling process, although quantum mechanically tunneling is 
a possible physical process. 
The partition goes back to Perelomov et al 
\cite{Perelomov:1966,Perelomov:1967I,Perelomov:1967II} (hereafter 
PPT model), it was based on the argument that the (semi-)classical 
trajectory is determined by the initial conditions 
$x(t_0)=0, v(t_0) = ik, k=({2 I_p})^{1/2}$, where $t_0$ is the time 
when the field reaches its maximum.  
The time was chosen such that the origin for the real time is the 
instant when $v_{exit}=v(t_R=0)=0$, i.e. $t_R=0$ when the particle 
emerges from the barrier, but at the same time \cite{Perelomov:1967II} 
the exit point is real, $x_{exit}=x(t_R=0)=x_c\ne 0$, where 
$x_c=I_p/F>> x_i$ is usually called the classical exit point. 
This makes the assumption that the T-time is a purely imaginary 
quantity  questionable.      
Tunneling is a quantum mechanical effect and to my best knowledge there 
exists no physical restriction, which forces us to assume that the 
dynamical time to overcome the barrier region, should be a purely  
imaginary quantity, which is a classical standpoint and did not reflect  
the quantum nature of the electron's motion in the barrier region, 
where certainly the energy conservation should not be violated, it 
turns out that a full understanding of the tunneling process in the 
ASEs is still waiting.

We note that in the PPT model, used by Torlina et al 
\cite{Torlina:2015}, one defines the inner region through the 
assumption that the approximate wave function $\psi_{in}$ can be 
substituted by the (field-free) ground-state wave function 
\cite{Kaushal:2013}. 
From this one sees that the border between the inner and outer regions 
in this model is comparable with the entrance point 
$x_{e,-}=(I_p-\delta_z)/(2F)\approx I_p/(2F)=\tilde{x}_e$ of the model 
of Kullie \cite{Kullie:2015,Kullie:2016} as seen in fig \ref{fig:ptc}.
For $F=F_a, x_{a}=x_{e,\pm}=Ip/2F_a$, this case (ionization 
as $\delta_z=0$) matches to the real part time as defined by 
Torlina, however $x_a$ differs from the classical exit point 
$x_{e,c}=I_p/F=2\, x_{a}$ assumed in Torlina model. 
As a side note, this means that $\tau_{T,i}=1/(2 I_p+\delta)$ can be 
chosen as an initial time $t_0=\tau_{T,i}$ for the Torlina model, 
see discussion below.
$\tilde{x}_e$ is also compatible with their choice of  the  radius
of the sphere separating the inner and outer regions 
$1/\kappa=(2 I_p)<<a<<Ip/F$ \cite{Torlina:2012I}. 
That means that the so-called under-the-barrier region as given in fig  
\ref{fig:ptc} (between $x_{e,-}$ and $x_{e,+}$ of the Kullie model) 
belongs roughly to the continuum (or outside region) of the Torlina 
model, although the model assumes that $x_{e,c}=I_p/F$ is roughly 
corresponds to the exit of the tunneling barrier. 
Hence, the tunneling process has a different meaning between the two 
models, where Torlina et al define it by an imaginary T-time, and by 
real and imaginary trajectories in the inside and outside regions, 
where ``tunnel-exit'' is a complex integrable singularity point of 
the potential.   
This can be compatible with the velocity gauge, where the barrier is 
absent, or not a physical one (the barrier region is crossed in an 
imaginary time elapse). 
Indeed, as mentioned above the definition of the barrier and the 
barrier regions in attosecond physics are gauge dependent 
\cite{Bauer:2005}. 
This suggests (but it needs a scrutiny to be accepted) that the 
Torlina model (when applied to the tunneling region $\gamma_K<1$) is 
quantitatively equivalent to the model of Kullie (provided both models 
have the same orders of approximation, see discussion below about the 
initial time), although the two models are completely different 
concepts, and the length gauge (used in Kullie model) has the advantage 
of presenting a clear physical picture \cite{BauerM:2016,Kullie:2016}, 
because no physical quantity corresponds to the vector potential in  
the velocity gauge. 
Note, Torlina et al also concluded that the total ionization rate 
depends only on what happens to the electron while it is tunneling 
\cite{Torlina:2013}, hence the confusion is mainly due to the 
different definitions of the tunneling process and the barrier region 
or the inside and outside regions. 

Another point is that the work of Torlina et al is an operative 
concept (compare below subsec \ref{ssec:dr}) it does not touch the 
concept of time (and its controversial discussion) 
in quantum mechanics. Indeed, it assumes implicitly a parametric time, 
point of view, compare eq \ref{Tnump}. 
It is worthwhile to mention that there exists classical procedures, 
which are quantitatively equivalent to the the Torlina model, e.g. 
the propagating of classical trajectories within a two-step model or 
(CCSFA) which involves classical Monte Carlo-type simulations 
\cite{Torlina:2013}. 
In the Kullie model there is no complicated calculation with a wave  
function, which is, no doubt, an advantage of the Torlina model. 
Also no trajectories (real or imaginary) are used, the calculations 
are achieved in time-energy space based on the TEUR. 
However, the geometry of the barrier is needed.  
It covers the time and T-time concepts in quantum mechanics, and makes 
use of, or offers a qualitative or quantitative connection to, the 
different fundamental issues of the quantum mechanics, i.e. the 
BE-photon-box-GE, the double slit experiment and the TEUR. 

Finally, the situation looks similar to the equivalence between my 
model and the FPI description of \cite{Landsman:2014II} as mentioned 
in subsec \ref{ssec:riq}.
At this point, one has to mention that a model called an entropic 
formulation of T-time \cite{Demir:2016} exists, the model is also based 
on the TEUR and the authors claim that their model describes a real 
T-time, which is in good agreement with Kullie model and the FPI model 
of \cite{Landsman:2014II}. 
However, crucial points still have to be clarified, e.g. choosing the 
reference system to calculate the delay time, where possibly it is one 
of the sources of the negative time in the Torlina model 
\cite{Torlina:2015} for $F> F_a$ ($F_a$ is the atomic field strength).
Indeed, one could consider it as an artifact effect, because Torlina 
et al evaluate their time starting at the peak of the spectrum, 
regardless the natural parameters of the system under consideration 
(the Hydrogen atom), i.e. the ionization potential (possibly with 
a Stark shift) and the atomic field strength, where the 
Above-threshold-ionization process starts.
The initial real time zero assumption $t_0=0$ at the peak intensity, 
in the attoclock of the Torlina model was criticized by Zimmerman 
et al, supplemental information \cite{Zimmerman:2016}. 
In my understanding, adding the term $\tau_{F_a}=1/(2I_p)$, see eqs 
\ref{Tnum}, \ref{Tnump}, to the Torlina model, can remedy the initial 
time problem (and the negative time), which enables one to take the 
system parameters in account and counts the delay (the real time part 
$t_R$ of $t_s$) relative to the ionization time $\tau_{F_a}$ at atomic 
filed strength. 
Using the initial time  $t_0=0$ at the peak of the spectrum
is equivalent to  eq \ref{Tnump} (compare fig \ref{fig:dnum1},   
\ref{fig:dnum2}), thus the real part of the calculated time by Torlina 
$t^{Torlina}_{R}=\tau_{num}$ or in other words, one obtains for the 
time of the Torlina model 
\[\vspace{-0.3cm}
Re(\tau_s)\equiv t_{R}^{Torlina}+t_0= \tau_{num}+t_0=\quad (t_0=\tau_{F_a})
\vspace*{0.0cm}\]
with $t_0=\tau_{d,F_a}=1/(2I_p)$. $Re(\tau_s)$ is then corresponds to 
the actual T-time $\tau_{T,d}$, see eq \ref{Tsym}. 
This becomes more apparent, when one compares the result of the 
numerical integration of the  TDSEQ (black squares) with $\tau_{T,d}$, 
$\tau_{num}$ in fig \ref{fig:dnum2}. 

Another reasonable possibility, to adjust the choosing of the initial 
time, is to consider $x_{e,-}$ as the initial position where the 
process (of tunneling or ionization) starts (which differs from the 
interaction instant, the first step see discussion in \cite{Kullie:2015}) 
and thus to consider the time to reach $x_{e,-}$ as the initial real 
time at the peak intensity, and hence $\tau_{T,i}=1/(2(I_p+\delta_z))
=t_0$, eq \ref{Tsym}, serves as a real initial time for the counting 
of the real time by the Torilna  model.   
The importance of this follows from the fact that only after the 
interaction \cite{Kullie:2015}, the propagation vector of the electron 
wave function is identical to the propagation vector of the optical 
vector potential. 
In addition, it is compatible, as discussed above, with the division 
inside ($\psi_{in}$) and outside of the barrier region at the point 
$x_{e,-}$ in connection to the Torlina approximation. 
In other words $\tau_{T,i}=t_0$ can be considered as the real part of 
the T-time  that correspond the imaginary part of the T-time of 
Torlina et al in the inside region, or the above mentioned expression 
of $t_s$ should be written in the form 
\[\vspace{-0.3cm}\tau_s=t_R^{Torlina} + t_0 + i\, t_T, \mbox{ with } t_0=
\tau_{T,i}=1/(2(I_p+\delta_z))
\vspace{-0.cm}\]  i.e. at the start (after the first 
step \cite{Kullie:2015}, the interaction step) the pointer of the 
attoclock points at $t_0=\tau_{T,i}$. 
At  $F=F_a$ this become $t_0=\tau_{F_a}=1/(2 I_p), \delta_z=0$, where 
the difference to $\tau_{T,i}$ is small because $\delta_z$ is small in 
the tunneling region.

Moreover, the imaginary part of the time is then the characteristic 
time of evolution of the wave function, in the sense brought by Messiah  
\cite{Messiah:1961} (see discussion in \cite{Kullie:2016}), and also 
could be present after the tunneling and not only inside the barrier. 

In the PPT model, used by Torlina et al and others, the time 
$t_s=t_R+i t_{T}$ or the solution of the saddle point equation, 
defines the starting point of the (semi-classical) trajectory 
\cite{Torlina:2013}. 
An imaginary time component to describe the T-time seems to be a 
result of the external time frame. A dynamical time should be 
connected to the system, i.e. one should use an internal clock. 
The solution of the saddle point equation is complex (i.e. of the form   
$Re(t)+ Im(t)$) but this did not means, that the T-time is the 
imaginary part of the solution and the real part is the ionization 
time after tunneling.  
In fact, Perelomov et al stated in their work \cite{Perelomov:1966} 
that the ionization time at atomic filed strength $F_a$ occurs in 
a time comparable with the atomic time, which is real and should 
serve as an initial point or the reference point ($\tau_{F_a}$ or 
$\tau_{T,i}$) of the attoclock, i.e. for the counting of the T-time or 
the ionization time in attosecond experiment, because at this field 
strength the barrier width equals zero (no barrier), and the barrier 
appears for $F<F_a$, as shown in our model, it increases gradually and 
becomes infinite for $F\rightarrow 0$, where the T-time becomes 
infinite \cite{Kullie:2015} (tunneling is not possible or the 
tunneling probability is zero) for the unperturbed system 
(ground state) as it should be.

Finally, unfortunately there is still no experimental data available for 
the Hydrogen (or a Hydrogen-like) atom \cite{Kielpinski:2014}, which 
is similar to the KASE for He, that makes it possible to compare 
the different time approximations  with an experimental finding and 
clarify (some of) the controversial issue of the T-time, especially 
the above mentioned points, keeping in mind that the two quantum 
mechanical concepts (of Kullie and Torlina et al) could be 
quantitatively equivalent or belong to two physically equivalent 
gauge pictures, similarly to other quantitatively equivalent treatments, 
such as the FPI treatment of \cite{Landsman:2014II} or the entropic 
real T-time of \cite{Demir:2016}.  
}
\vspace*{-0.50cm}
\subsection{\small{The different regimes}}\label{ssec:dr} 
\vspace*{-0.2cm}
Despite the different views concerning the tunneling process, it is 
still puzzling and rather not well understood. 
In the SFA one usually assumes a photoelectric effect mechanism or 
multiphoton absorption (multiphoton processing), it is important in 
the regime of large Keldysh parameter $\gamma_k>1$ and also is the 
usual process in quantum optics experiments, although in quantum 
optics (where the field strength is weak) usually an operational 
point of view is used \cite{Scully:1998,Muthukrishnan:2003}, based 
on the second quantization formalism, for the theoretical treatments
and understanding of the experimental findings. 
It is initiated by  Rony Glauber \cite{Muthukrishnan:2003}
[A photon is what a photodetector detects], which is in line with 
the operational time-of-arrival concept mentioned above. 

Taking into consideration the view of Collins \cite{Collins:1987} as 
discussed above, we prefer, see sec \ref{sec:TP}, another point of view 
in the region of small Keldysh parameter $\gamma_k<1$ (the tunneling 
regime), and argue that a scattering mechanism (or elastic collision) 
is involved in the tunneling process in attosecond experiments such 
KASE, where a large number of photon are involved, which means that 
the electron recoil due to the scattering with LWP, with a drift 
along the radiation direction \cite{Meyerhofer:1997,Moore:1995}. 
It is worthwhile to mention that even the Compton effect can be 
explained by a semiclassical nonrelativistic approximation, 
Schr\"odinger \cite{Schroedinger:1927}, see \cite{Kuhn:1995} p. 
222-225, where the electron is described by a wave function. 
Indeed, the experimental investigation of 
\cite{Meyerhofer:1997,Moore:1995} showed a type of nonlinear 
Compton scattering at high laser intensities, where many photons  
participate in a single scattering. 
Theoretically, an earlier work of Eberly \cite{Eberly:1965} proposed 
an experiment with high photon density $\rho_L (Watt/cm^3)$ in the 
optical frequency range $\omega$, where  the effect to be observed is 
that the photons of the laser beam interact  (collectively) 
simultaneously with the electron, and give up momentum and energy to 
the electron depending on a nonlinearity dimensionless parameter  
$0<\varepsilon\sim \rho_L/\omega<1$. 

This is important because it also brings up the particulate nature 
of the radiation in the discussion, although Lamb, in a nice paper 
titled ``anti-photon'' \cite{Lamb:1995}, concluded and argued to give 
up the photon as a particle [It is high time to give up the use of 
word ``photon'', and of a bad concept which will shortly be 
a century old. 
Radiation does not consist of particles, and the classical, i.e. 
non-quantum, limit of quantum theory of radiation is described by 
Maxwell's equations for the electromagnetic fields, which do 
not involve particles.] 
Ironically, Einstein himself did not accept wave-particle 
duality (WPD). He wrote: this interpretation... (WPD) .. appears to me 
as only a temporary way out \cite{Schilpp:1949}. 
Although, in the earlier days of quantum mechanics, Duane 
\cite{Duane:1923} was able to explain diffraction of x-rays by 
a crystal in terms solely of their particle aspect. 

However, modern experiments of quantum optics have beautifully 
confirmed the corpuscular character of the photon, 
Zeilinger \cite{Zeilinger:2007,Zeilinger:2005}. 
Furthermore,  Zeilinger concluded that the results suggest that one 
has to abandon at least the notion of na\"ive realism that particles 
have certain properties that are independent of any observation 
(pre-existing properties). 
We will rely on this conclusion of Zeilinger and see that the KASE, 
is a case (and generally in the ASEs) to realize the existence of 
a corpuscular property of the (attosecond pulse) radiation, precisely  
of an attosecond LWP. 
\vspace*{-0.25cm}
\section{The tunneling process}\label{sec:TP}
\vspace{-0.25cm}
A hallmark of the tunneling regime in strong field ionization is 
the Keldysh parameter $\gamma_K$, eq \ref{gamK}. 
For $\gamma_K<1$ it follows that
\begin{equation}\label{gamk}
 \gamma_K=\frac{\sqrt{2 I_p}}{F}\omega_0< 1 \Rightarrow 
 (\frac{F}{\omega_0})> \sqrt{2 I_p}=k_e
\end{equation}
where $k_e$ is the wave vector or the momentum of the electron. 
The left hand side of the second inequality of the relation 
(\ref{gamk}) can be interpreted  as an average momentum of the laser 
pulse $p_{L}=k_L$. This leads to the following:
\begin{equation}\label{mL}
(\frac{F}{\omega_0})=k_L\, =\, M_L\, c,\mbox{ with } 
 M_L\, =\,(\frac{k_L}{c}) =\, (\frac{F}{c\, \omega_0}) 
\end{equation}
where $M_L$ is an "effective'' average mass (or simply 
effective mass), a feature of the whole photonic laser pulse or the LWP, 
$c$ is the speed of light and $\omega_0$ (in $au$) is the photon energy 
or the central circular frequency of the LWP. $M_L$ has nothing to do 
with the rest mass of a single-photon, which has an upper limit of 
$m_{ph}\le 10^{-49} g \approx 10^{-17} eV$ \cite{Tu:2005}. 
Indeed, different authors suggested a mass characteristic for the light  
pulse, where experimentally it was found that the light pulses 
propagate in the vacuum with a speed somewhat smaller than the speed of  
light $c$ \cite{Giovannini:2105}.  
Fedorov \cite{Fedorov:2016} claims that a nonzero invariant mass of 
\begin{equation}\label{mFed}
m_{Fedorov}=N\frac{\lambda_0}{2\pi w_a}\frac{\hbar\omega_0}{c^2}
\end{equation} 
can be attributed to the light pulse, where $w_a>>\lambda_0$ is the 
pulse width (waist) at $z= 0$, at the center where a Gaussian pulse 
becomes approximately a plane wave, $\omega_0=c k_0$  the central 
frequency of the laser pulse  and $N$ the photon number.  
$m_{Fedorov}$ characterizes global features of the pulse as a whole. 
It is astonishing to see that for $\lambda_0=2 \pi w_a$ one gets 
$m_{Fedorov}/N=m_{ph}=E/c^2=\hbar\omega_0/c^2$.   
The invariant mass of pulses is shown to be related directly with the 
propagation velocity of pulses in vacuum, which is found to be smaller
than the light speed in all cases except an infinitely extended plane 
wave \cite{Tu:2005,Fedorov:2016}.

The effective mass ${M_L}$ differs from the ``sum'' of all the 
single-photon masses $\overline n_{ph} \cdot m_{ph}=\overline  n\, 
\frac{\omega_0}{c^2}$, where $\overline n_{ph}$ is the mean number of 
the photon in the pulse, in this case $m_{ph}$ is obtained form the 
relation $E=m_{ph}\,c^2$ according to the theory of relativity. 
$m_{ph}=E/c^2=\hbar \omega_0/c^2$ has been introduced formally as an 
effective mass entering a formal Schr\"odinger equation for a 
``photon-wave function'', the field amplitude $F$ \cite{Kuhn:1995} 
page 153. 
It is clear that a wave packet is not a single-photon wave and only in 
the limit of a plane wave, is $\omega_0$ the energy carrier of 
a single photon  (where the photons density related to the intensity 
$I_L\sim F^2$ via $I_L/\omega_0$.)  
We can consider the LWP as a carrier of a finite number of energy 
quanta, the (average) number of photons in the pulse. 
Similarly, on the particle side a plan matter-wave corresponds to 
a particle stream and a single particle is described by a matter-wave 
packet (MWP). 
At this point I have to stress not to anticipate any  possible 
misinterpretation of my claim in this section. 
There is no reason to interpret $M_L$ (or $m_{ph}$) as a mass of 
a small bullet (point-particle) or a group of point-particles. 
However, light quanta or the photon can be considered as a particle 
in the sense that it shows particle behavior as claimed by 
Grangier et al \cite{Grangier:1986}.
And according to Compton \cite{Compton:1924}  
(see \cite{Kuhn:1995}, page 224): one  concludes light can also 
consist of discrete units moving in certain directions, each unit 
has the energy $\hbar\, \omega$ and the impulse $\hbar\, \omega/c$. 
Hence for a radiation of LWP  with an approximate intensity 
$I_L \approx \omega_0=\omega_{ph}, n_{ph}\approx1$, one obtains   
\begin{eqnarray}\label{ke}
&&\lim_{I_L\approx\omega_0}{{M_L}\approx m_{ph}}= \omega_{ph}/c^2
\Rightarrow  F_{ph}={\omega_{ph}^2}/{c}={E_{ph}^2}/{c}
\end{eqnarray}
Although it is difficult to interpret this relation, but a possible 
simple interpretation would be, $F_{ph}$ is a limit of the electric 
field strength (for a fixed circular frequency $\omega_0$) for 
which a LWP can approximately be considered as a single-photon pulse, 
in the sense that it can show particle behavior. 
A similar situation when 
the De Broglie wave length becomes smaller than the geometry of the 
particle. 
For the experimental setup of the attosecond experiment, in the 
optical range, this is a very small value ($F \sim 10^{-5} au$), 
the typical field strengths used in the ASEs $\sim 0.01-0.11 au$.
In the context of the field quantization, according to Purcell 
\cite{Purcell:1956} (see \cite{Kuhn:1995}, page 153), it makes no 
difference whether we think of $\rho=\widehat{F}^2$ as a square of 
the electric field strength or the photon probability density $\rho$.

Now we can turn back to eq \ref{gamk} keeping in mind the above 
discussion, especially that our concern is the impact of the LWP as 
a carrier of momentum, when it interacts with the electron, where LWP 
is a group of light quanta or photons propagating in the vacuum with 
a speed $v_{c}$ slightly smaller than the speed of light $c$ 
\cite{Giovannini:2105,Fedorov:2016}, and showing a particle-behavior. 
Any further interpretation is beyond the scope of this work. 
Eq \ref{gamk} has a simple interpretation in the language of QM, for 
${k_L}< k_e, \, (\gamma_K>1)$ the ionization happens through 
multiphoton absorption (MPI regime). 
Whereas for ${k_L}> k_e, \, (\gamma_K<1)$ the tunneling regime, 
one figures out that, as the average LWP momentum is larger 
than the electron momentum, the tunnel ionization happens by a heavy 
scattering of the (bound) electron through the LWP. 
The LWP acts collectively and strongly and the electron is therefore,   
not able to form immediate metastable or virtual states, as it is 
the case in the MPI regime. 
The electron response happens on a fast time scale. Consequently, and 
unlike the MPI regime,  the ionization depends mainly on the field 
strength, and not significantly on the frequency (which is significant 
for the MPI) \cite{Delone:2000} chap. 1, and the electron rather 
follows the value of the laser intensity 
(the intensity envelope of LWP) \cite{Delone:2000} chap. 2.3, 
see below, thus the electron undergoes a dynamical transport 
\cite{Collins:1987}, see sec \ref{ssec:mda}.     
Conceivably, which is justified by Collins \cite{Collins:1987}, the 
scattering time concept (the decay of metastable states, compare 
subsec \ref{ssec:mda}),  when used to calculate the T-time in the 
ASEs, has no success in cases such as KASE as shown by Landsman 
\cite{Landsman:2015,Landsman:2014II} and already mentioned in 
subsec \ref{ssec:mda}.  

In accordance with the view of Colllins, subsec \ref{ssec:mda},  
we suggest a scattering mechanism (or elastic collision) in the 
tunneling regime as mentioned above subsec \ref{ssec:dr}, it is not 
that of a single photon scattering (Compton scattering),  but that the 
LWP collectively scatters the electron.  
This collective process shows a particulate nature of the LWP, 
similar to the way the collective process of the particles shows 
the wave behavior on the screen 
in the double slit experiment (DSE). 
The latter happens even when one sends the electron one-by-one 
\cite{Merli:1976},\cite{Rosa:2012} (voted the most beautiful 
experiment by readers of Physics World in 2002, Sep 1.)  
For the former similar experiments can be performed, like 
the single-photon double-slit experiment \cite{Grangier:1986}, 
or the Compton scattering experiment 
\cite{Compton:1923,Compton:1923I} with single-photon LWP provided 
is experimentally achievable.   
Zeilinger \cite{Zeilinger:2005} concluded that the results of these 
(achieved) experiments confirm that the quantum state is not just 
a statistical property of an ensemble of particles. 
The conceptual questions arising for photon interference are 
the same as those arising for interference of massive particles,  
and in both cases we see particle-like and wave-like properties.  
And that inequivalence for certain interference experiments 
arises  because the photon has no rest mass \cite{Zeilinger:2005}. 
Clearly, the De Broglie wave relation $\lambda=1/p$, with $p$ 
the momentum, is the basis for the wave nature of a particle and 
the basis for the photon hypothesis as a carrier of a momentum 
$k=m_{ph} c=\omega/c$ \cite{DeBroglie:1929} 
(De Broglie Nobel-prize lecture), see eq \ref{DB} below, where $m_{ph}$ 
here is usually interpolated as the ``motional'' mass of the photon 
\cite{Compton:1923}.
However, one has to avoid a misinterpretation of $m_{ph}$ because the 
rest mass of the photon is zero ($<10^{-49} g$) \cite{Tu:2005}.   

In table \ref{tab:ML} the effective mass $M_L$ for a range of 
intensities used in the ASEs is given. 
The average number of photons {$\overline{n_{ph}}$ is calculated with 
the assumption that the laser pulse hits the electron in its ground 
state orbital ($1s$ for He atom, with the radius $r_e \approx 1 au$), 
with the cross-section $A_e= 4 \pi r_e^2=4 \pi$, the area of a sphere 
of the ground state during the time period of one cycle 
$\delta t_e=1/\sqrt{2I_p}$ (i.e. such that the probability 
amplitude of finding the electron on a spherical shell of the ground 
state during the time $\delta t$ is $\psi^2=1$.)   
The average number of photons is then 
$\overline{n_{ph}}=(I_L/\omega_0) A_{e} \delta t_e$. 
Assuming a photon Compton scattering, see \cite{Compton:1923} and 
\cite{Compton:1923I}, the momentum transmitted to the electron is 
$\sim(\frac{\omega_0}{c})$ \cite{Compton:1924}, which is too small 
(as expected) relative to the momentum of the electron $k_e$, 
$\frac{\overline{p_{_C}}}{k_e}$ is given in percent in table 
\ref{tab:ML},  where 
$\overline{p_{_C}}=\overline{n_{ph}}(\frac{\omega_0}{c})$.  
Note that a semiclassical approximation due to Schr\"odinger 
\cite{Schroedinger:1927} explains the Compton effect. 
Whereas assuming a collective LWP scattering, the average momentum 
transmitted to the electron is $\overline{p_{W}}$. 
The quantity $\frac{\overline{p_{W}}}{k_e}$, given in table 
\ref{tab:ML} in percent relative to the momentum of the electron, 
is calculated as the following, see \ref{ap:WK}:
\begin{equation}\label{pw}
\overline{p_{W}} = \, \eta \, \, \alpha \,
\left(\frac{F}{\omega_0}\right)^2, 
\end{equation}
where $\alpha=1/c$ is the fine structure constant, which is 
equal to the strength of the interaction of the photon with 
the electron.  
Note that the interaction of an electron with an intense laser field 
is characterized by $(F/\omega_0)^2$ \cite{Meyerhofer:1997} 
(and ref 10 inside it.)   

In eq  \ref{pw} the quantity $\overline{p_W}$  
determines the amount of an average momentum that is transmitted to 
the electron, and is (depending on the  unknown parameter $\eta$) 
much larger than $\overline{p_{_C}}$ of  Compton 
scattering as seen in table \ref{tab:ML}. 
One notes that in the above consideration 
$\overline{p_W}/\overline{p_C}$  
$\sim \overline{n_{ph}}^{-1}\, (F^2/\omega_0^3)=
\eta\, (\delta t\, A_e \,\omega_0^2\, (1+\epsilon^2)^{-1} 
= \eta\,\cdot 15.89$ ($\omega_0,\,\epsilon$ are given in 
sec \ref{sec:In}). 
It is worthwhile to mention that in strong field experiments, the 
process of scattering in the tunneling and the MPI regimes are complex 
and nonlinear \cite{Delone:2000}. 
As the process is nonlinear, it is possible that $\eta$ depends on 
the mean photon number $\overline{n_{ph}}$ involved in the interaction 
(not the total number of the photons of the LWP), which depends on 
the probability density to find a photon in the interaction 
volume/area, compare $m_{Federov}$ eq \ref{mFed}, which depends on $N$.   

Hence the determination of a precise value of $\eta$ needs 
further investigations, perhaps using wave packet dynamics.
Indeed, it will also be fruitful to achieve a Compton-type experiment 
using strong field attosecond isolated short pulses, e.g. LWP in the 
visible light, UV or XUV range.
Today the experimentally achievable pulse durations are  sufficiently 
short, about $67-130\, as$ \cite{Zhao:2012,Kling:2008}. 
Note eq \ref{pw} is related to $U_p/c$, where $U_p$ is the well 
known Ponderomotive energy and is defined as the cycle-averaged 
quiver energy of the electron in the laser field. 
For a free electron $U_p={\chi}\, F^2\, \omega_0^{-2}$, $\chi=1/2$ 
(for linear polarized laser field).  
\begin{table}[t]
\scriptsize
\begin{tabular}{lllllllll}\hline
$x^a$&1& 2&3& 4& 5& 6& 7.5\\\hline
$I_n (au)$& 0.00285& 0.0057&0.00855& 0.0114& 0.01425& 0.0171& 0.0214\\  
$F (au)$  & 0.040  & 0.057 & 0.07  & 0.081 & 0.090  & 0.099 & 0.11\\
${M_L}$   & 0.0048 & 0.0067& 0.008 & 0.0095& 0.0106 & 0.0116& 0.013\\
$\overline{n_{ph}}$& 0.43  & 0.86  & 1.3& 1.72&2.15& 2.6 & 3.23\\
$\frac{\overline{p_C}}{k_e}\%$ &  0.0145 &  0.023 & 0.043 & 0.058 & 0.072 
& 0.087 & 0.11 \\
$\frac{p_{W}}{k_e}\%$ &0.23& 0.46& 0.69& 0.92& 1.15& 1.38& 1.72\\\hline
\end{tabular}
\caption{\label{tab:ML}\scriptsize
$^a$ Intensity$=x \ast 10^{14}$ in $W cm^{-2}$, $I_n=$ intensity, 
$F=$ field strength, $\overline{n_{ph}}=$ mean photonic number,
$M_L=\alpha\, (\frac{F}{\omega_0})$, 
$\overline{p_C}=\overline{n_{ph}}\, \alpha{\omega_0}$, 
$\overline{p_W}=\eta\, \alpha\, F^2/\omega_0^2$, 
$\eta =1$, see text.}
\vspace*{-0.5cm}
\end{table}

In MPI regime the average photon momentum is not large enough to 
destabilize the electron in its circular movement, 
and the electron is captured in a metastable state (by absorbing 
a photon) with a new wave vector $k^{'}_e<k_e$ and a new oscillation 
frequency $(e.g.\, \omega_{e}^{'}=\omega_{e}-\omega_0<\omega_{e}$). 
In the case of virtual states  the electron evolves successively 
through the virtual states to a final orbital or to the continuum) 
by absorbing portions of  energy (many photons) from the radiation.   
Einstein's insight to the Planck hypotheses was to see that absorbing 
energy from radiation in a quantized form leads to the fact 
that light implies a quanta (photon), which carries the energy 
$\omega_0$, although the MPI process, and the photoelectric effect 
itself, is not the regime, where the corpuscular property emerges 
from the radiation or from a laser pulse, i.e. no need for the 
quantization of the classical electromagnetic  radiation 
(only quantization of the matter) \cite{Grynberg:2010}. 
Similarly the Compton effect can be explained by a semiclassical 
approximation as mentioned above  \cite{Schroedinger:1927}. 
In the attosecond experiment we see the advantage that in 
the same experiment, both properties of the radiation can emerge,  
the  wave property (in MPI regime, no photon-particle concept needed) 
and the corpuscular property (having particle-behavior) in the 
tunneling regime.
Roughly the range of $\gamma_K$ determines the different regimes.  
Likewise the DSE with light (Young experiment) is used to prove 
the wave nature of light and the DSE with matter is used to prove 
the wave nature of particles.  
That the KASE (in the tunneling regime) is also a DSE, was discussed 
in \cite{Kullie:2016} in accordance with the use of the Feynman path 
integral by Landsman \cite{Landsman:2014II} to calculate the T-time. 
Hence, the wave/particle duality nature of the radiation can be 
impressively shown in the ASEs such as the KASE, with the caution that 
the present work presents only a step in this direction, and certainly,  
the experiment will enlighten and is decisively important to prove 
our approach.
We discuss this further by drawing the following remarks to our 
attention.  
\begin{itemize}
 \vspace{-0.3cm}
\item The wave packet concept represents an unifying 
 mathematical tool that can cope with and embody nature's 
 particle-like behavior and also its wave-like behavior,  
 Zettili \cite{Zettili:2009}, chap 1.8.
 \vspace{-0.3cm}
\item ${M_L}$ is  an effective mass of the LWP, which acts 
collectively (like a MWP) at a time scale much smaller 
 than $1/\omega_0$.  
  \vspace{-0.3cm}
\item The corpuscular property can be best judged form the momentum, 
$p_{photon}, p_{particle} =\hbar k$. 
For a stream of photons or a plane wave (the ``motional'' mass of 
a photon $ m_{ph}=\omega_0/c^2$) $\overline{m_{ph}} \sim 
\overline{n_{ph}}\, m_{ph}= \overline{n_{ph}}\, \omega_0/c^2$,
where $\overline{n_{ph}}$ is the average number of the photons.
For a LWP $M_L={k_L}/c= \frac{F}{c\,\omega_0}$
 \vspace{-0.3cm}
\item From  eq \ref{gamk},  
${k_L}>k_e$ or $c \, {M_L}>k_e$,  where  ${M_L} ={k_L}/c$, 
the corpuscular nature (particle-behavior) of the  LWP is a reasonable 
assumption, because the De Broglie relation is based on the symmetry 
between light (wave-light) and particles, as is clear in the following 
relation \cite{DeBroglie:1929}
\begin{equation}\label{DB}
\left\{
\begin{array}{ll}  
m_{ph}\, c=\frac{E}{c}=\frac{\omega_{ph}}{c}={p}&
\mbox{the particle nature}\\
E=c\,k , \lambda_D=\frac{1}{p}&    \mbox{the wave nature}
\end{array}
\right.
\end{equation}
 \vspace{-0.cm}
\end{itemize}
De Broglie wrote:  
I was guided by the aim to perform a real physical synthesis, 
valid for all particles, of the coexistence of the wave and of 
the corpuscular aspects that Einstein had introduced for photons 
in his theory of light quanta in 1905 \cite{DeBroglie:1970}.
Naturally the symmetry in eq \ref{DB} permits both  directions, wave 
property for particles or MWP and a corpuscular property of the 
radiation or LWP. 
Indeed, De Broglie insight was to see that for the light quanta from 
$E=h \nu= h c/\lambda$ and $E=c p$ it follows that $\lambda= h/p$ and 
the same holds for the particles, where $h$ is the Planck constant. 

Today the question is clear: under which conditions of an 
experimental setup or an observation, is a property reflected. 
Clearly this point of view did not contradict the operational 
point of view, which circumvents the natural (internal) process of 
detecting the photon. But additionally, the former considers  the fact 
that an instrumental observation, measuring quantities, reflects 
a certain property as mentioned above 
\cite{Zeilinger:2005,Zeilinger:2007}, see sec \ref{ssec:dr} 
last paragraph.    
The famous DSE shows the wave nature of the particles, the modern 
quantum optics experiments show the particulate nature of the 
radiation, and similarly in the KASE-type experiment we encounter 
the corpuscular property of LWP. 
In a related issue, as mentioned above, Fedorov \cite{Fedorov:2016} 
showed that the invariant mass of pulses (LWP) eq \ref{mFed}, 
characterize a global feature of the pulse and is related directly 
with the propagation velocity of pulses in the vacuum.  
In the LWP, it is easy to figure out that the average momentum is 
a collective effect ($M_L$)  and is equivalent to the DSE for 
particles, in which the collective effect of the particles shows the 
interference (wave-property) picture on the screen. 
Hence the question is, when using a ``single-photon'' LWP,
would it show the same effect (for $\gamma_K<1$), and how similar  
is it to the Compton effect \cite{Compton:1923}.  
It is instructive to mention at this point that in 1986 Grangier 
et al \cite{Grangier:1986} reported a modern laser based version of 
Taylor’s experiment \cite{Taylor:1909}. 
They provide convincing evidence, that with a suitable care one can 
prepare single-photon states of light. 
Such photon states, when sent to a beam splitter, display the type of 
statistical correlations we would expect of particles. 
In particular the single photon appears to go one way or the other. 
Yet such single-photon states can interfere with themselves, even when 
run in ``delayed choice'' \cite{Zajonc:2003,Hellmuth:1987}, in 
a way similar to the particles in the DSE \cite{Kuhn:1995} p. 213.

To conclude, the situation in the attosecond experiment, such as the 
KASE, in which the collective act of the LWP shows (for $\gamma_K<1$) 
a particulate property $k_L, M_L$, is similar to the DSE, in which the 
collective act of the particles shows a wave property, even when the 
particles are sent to the screen one-by-one 
\cite{Zeilinger:2005,Merli:1976}. 
And similarly, in the case of a particulate property for photons 
\cite{Grangier:1986,Hellmuth:1987} (see also 
\cite{Tippler:1969,Kuhn:1995} p. 153), the interference depends on 
the number of the photons, even when the field strength is so small 
that the photons reach the screen only one-by-one.

In the MPI region ($\gamma_K>1$), the evolution of the electron wave 
packet is under intensive research. Unfortunately, the measurements 
are restricted to the relative delays between two photo-electron wave 
packets ionized from two different orbitals, for details the reader 
is kindly referred to the tutorial \cite{Dahlstrom:2012}.
One notes that the Keldysh time $\tau_K$ is large, for in the region 
of $\gamma_K>1$ the field strength of the laser pulse is small and 
does not disturb the electron heavily, the evolution of the electron's 
MWP is relatively slow, while for $\gamma_K<1$  the electron is 
enforced to move at a fast time scale $\tau_{T,sym}\, 
(\mbox{or } \tau_{T,d})$, see eq \ref{Tsym}. 
Furthermore, in the immediate region $\gamma_K\sim 1$, which is 
narrow \cite{Delone:2000} (chap 1, p. 2), there is still no clear 
picture, where both the MPI and tunneling can take place 
\cite{Ivanov:2005,Klaiber:2015} and investigations show that 
non-adiabatic effects affect the tunneling, i.e. differently from 
the adiabatic approximation, where the electron sees approximately 
a static electric field during the ionization process 
\cite{IAIvanov:2014,Geng:2014,Li:2016,Hofmann:2014,Hofmann:2016}.   
\paragraph*{\scriptsize \bf Conclusion}
We have discussed in this work different points in the ASEs, related to 
the issue of the tunneling time, the tunneling process and wave-matter 
duality. 
The tunneling time $\tau_{T,d}$ differs from the parametric time, 
which equals the numerical $\tau_{num}$, which is obtained by the 
numerical integration of the TDSEQ.  
We proposed a scattering mechanism of the electron by the laser wave 
packet in the region of the Keldysh parameter $\gamma_K<1 $, and 
argued that the corpuscular nature (particle-behavior) of radiation 
emerges in this region, similar to the double slit experiments in 
quantum optics, where the wave-matter duality of radiation and 
particles is proved. 
\tiny
\vspace*{-0.50cm}
\appendix
\section{}\label{ap:WK}
\vspace*{-0.25cm}
Due to the scattering process with the electron, the LWP with an 
average momentum $k_L$ undergoes a perturbation. We can expand the 
momentum $k_L^{\prime}$ in powers of $(\frac{F}{\omega_0})=k_L$, where  
$k_L^{\prime}$ is the average momentum of the LWP  after it scatters the 
electron ($F,\, \omega_0$ as in sec. \ref{sec:TP}),
{\tiny
\begin{equation}\label{ap:wk}
k_L^{\prime}=p_L^{\prime}({\omega_0,F})  =  \sum_{i=0}\, a_i k^{(i)}=
a_0\, k^{(0)} + a_1\, k^{(1)}  + \cdots   
\end{equation}}
\noindent
The perturbation terms in \ref{ap:wk} obey,
{\tiny
\begin{equation} 
k^{(i)}=\alpha^{i}\, \left(\frac{F}{\omega_0}\right)^{(i+1)},\,  
i=0,1,2 \cdots
\end{equation}}
\noindent
where $\alpha={1}/{c}$ (the fine structure constant), $c$ the speed of light.
Without loss of generality, we can take $a_0$=1, and 
{\tiny 
\begin{equation}\label{ap:WK1}
a_{i\ge1}= \left\{
\begin{array}{ll}  
0     &\mbox{ for } k_L=p_L({\omega_0,F}) \\
\ne 0 &    \mbox{ for }k_L^{\prime}=p_L^{\prime}({\omega_0,F})
\end{array}
\right.
\end{equation}}
\noindent
i.e. for the initial wave packet we set $k_L= k^{(0)}
=(\frac{F}{\omega_0})$.
The series can safely be truncated to the first order in $\alpha$ 
(second order in $(\frac{F}{\omega_0})$). 
The small change in LWP  is of the order 
$ k_L- k_L^{\prime}  \sim  \alpha \left(\frac{F}{\omega_0}\right)^2$,
and is about $1\%$ of $k^{(0)}$. This finally leads to eq \ref{pw}.

It is worthwhile to mention that, Meyerhofer \cite{Meyerhofer:1997}, 
the traditional boundaries between Thomson and Compton scattering 
become less clear. 
For example, nonlinear Compton scattering of photons and an electron 
takes place in the presence of an intense laser field, mixing  quantum 
mechanical and classical pictures.
Hence eq \ref{pw}, sec. \ref{sec:TP}, is valid for LWP scattering,  
where $\eta$ depends on the mean number of the photons involved in the 
interaction, not on the total number of the photons of the pulse $N$.
\subparagraph{\scriptsize \bf Acknowledgments}
I would like to thank Prof. Martin Garcia from the Theoretical 
Physics of the Institute of Physics at the University of Kassel 
for his kind support.
\tiny
%
\end{document}